\documentclass[10pt,a4paper,twocolumn]{IEEEtran}

\usepackage{amsmath,amssymb,verbatim,cite,amsopn,graphicx,citesort,url,color}
\newcommand{\vr}{\gamma}

\begin{document}

\title{Wireless Physical Layer Security with Imperfect Channel State Information: A Survey}

\author{\IEEEauthorblockN{Biao He, Xiangyun Zhou, and Thushara D. Abhayapala}
\thanks{B. He, X. Zhou and T. D. Abhayapala are with the Research School of Engineering, the Australian National University, Australia (e-mail: {biao.he, xiangyun.zhou, thushara.abhayapala}@anu.edu.au).}
}

\maketitle

\begin{abstract}
Physical layer security is an emerging technique to improve the wireless communication security, which is widely regarded as a complement to cryptographic technologies. To design physical layer security techniques under practical scenarios, the uncertainty and imperfections in the channel knowledge need to be taken into consideration. This paper provides a survey of recent research and development in physical layer security considering the imperfect channel state information (CSI) at communication nodes.  We first present an overview of the main information-theoretic measures of the secrecy performance with imperfect CSI. Then, we describe several signal processing enhancements in secure transmission designs, such as secure on-off transmission, beamforming with artificial noise, and secure communication assisted by relay nodes or in cognitive radio systems. The recent studies of physical layer security in large-scale decentralized wireless networks are also summarized. Finally, the open problems for the on-going and future research are discussed.
\end{abstract}


\section{Introduction} \label{sec:Intro}
Secure communication over wireless fading channels becomes a critical issue due to the broadcast nature of wireless networks.
Traditionally, key-based cryptographic technologies~\cite{Massey_88} are used to secure the data transmission.
However, the secrecy provided by cryptographic technologies is conditioned on the premise that the eavesdroppers have limited computational capability to decipher the message without the knowledge of secret keys. This premise becomes controversial with the rapid developments of computing devices.
On the other hand, physical layer security is an emerging research area that explores the possibility of achieving perfect-secrecy data transmission among legitimate network nodes, while possible malicious nodes eavesdropping the communication obtain zero information~\cite{Bloch_08}.
Basically, the objective of physical layer security is to minimize the amount of confidential information that can be obtained by the illegitimate users according to their received signals. To achieve secure communications over wireless channels, physical layer security explores time-varying properties of the fading channel, smartly designs the channel code, and processes the transmitted signals, instead of relying on encryption.

\begin{figure}[!b]
\centering\vspace{-2mm}
\includegraphics[width=1\columnwidth]{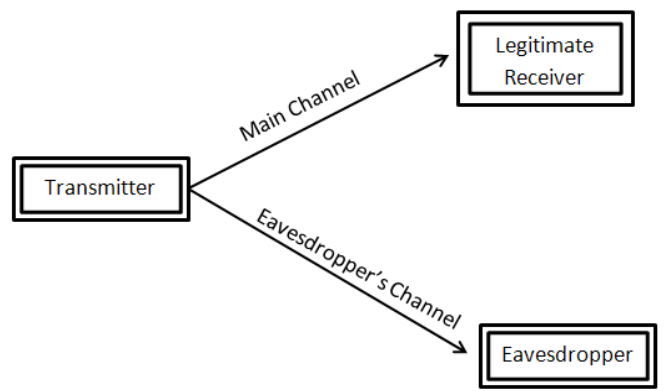}
\vspace{-20mm} \caption{Basic system model of physical layer security over wireless channels.} \label{fig:Fig1}
\end{figure}

The information-theoretic foundation of secret communication was laid by Shannon in~\cite{Shannon_49}.
Wyner's pioneering work introduced the wiretap channel model as a basic framework for physical layer security~\cite{wyner_75}, which was extended to broadcast channels with confidential messages described by Csisz´ar and K¨orner in~\cite{csiszar_78}. These early works have led to a significant amount of recent research activities taking the fading characteristics of wireless channels into account. The basic system model of physical layer security over wireless channels is shown in Fig.~\ref{fig:Fig1}.
Most works in this area rely on the perfect knowledge of both the legitimate receiver's and eavesdropper's channels at the transmitter to enable secure encoding and advanced signaling.
However, the assumption of perfect knowing channel state information (CSI) is not realistic.
In practical scenarios, many reasons can cause imperfections in the CSI at the communication nodes, such as the following ones:
\begin{itemize}
  \item No feedback from the eavesdroppers. When the eavesdropper is a passive entity, its CSI or even location is very hard to be obtained at the legitimate transmitter. Also, when the eavesdroppers are malicious users, they are not willing to provide their channel information to the legitimate party.
  \item Partial CSI feedback from the receivers. The receivers sometimes can only provide partial CSI feedback to the transmitter, e.g., limited-rate feedback, channel direction feedback, and signal-to-noise ratio (SNR) feedback.
  \item Imperfect feedback links between the transmitter and receivers.  When the feedback links are not error-free or delayless, noise component is added into the feedback information, or the CSI obtained at the transmitter is an outdated version of the actual CSI.
  \item Channel estimation errors at the receivers. Since the estimation of fading channels generally is not error-free, the CSI obtained at the receivers is not perfect.
\end{itemize}
Over the past few years, increasing attention has been paid to the impact of the uncertainty in the CSI on both legitimate receiver's and eavesdropper's channels. The remainder of this paper is devoted to surveying and reviewing the literature of physical layer security with imperfect CSI\footnote{Note that the work considering systems with imperfect instantaneous CSI often covers the case of system with no instantaneous CSI, since imperfect instantaneous CSI with very large uncertainty naturally converts to the case of no instantaneous CSI.} in wireless communications.
We aim to provide a high-level overview of the current research and development of the field.
In addition, this survey focus on the physical layer security researches without using the secret key, despite that some physical layer security work,  e.g.,~\cite{Maurer_03,Maurer_03_2,Chou_09,Lai_12,Lai_12_2}, investigated the secure transmission with the key observable by the eavesdropper over wireless channels.

The remainder of this paper is organized as follows.
For research from the information theoretic perspective, Section~\ref{sec:IT} presents the main performance metrics of secure transmissions with imperfect CSI. Signal processing secrecy enhancements developed in the secure transmission design considering imperfect CSI are reviewed in Section~\ref{sec:SP}. Section~\ref{sec:LargeSc} turns to the secrecy with channel uncertainty in large-scale decentralized wireless. The open problem and possible future research directions are described in Section~\ref{sec:OP}. Finally, Section~\ref{sec:Conc} concludes the paper.

\section{Characterizations of the performance limits}\label{sec:IT}
The performance limits of secure transmission systems with full CSI are often characterized by the secrecy capacity.
The secrecy capacity, $C_s$, for degraded wiretap channel with additive Gaussian noise was given by~\cite{Cheong_78},
\begin{equation}\label{}
  C_S = C_M - C_E,
\end{equation}
where $C_M$ and $C_E$ denote the Shannon capacities of main (legitimate receiver's) and eavesdropper's channels, respectively. A positive secrecy capacity can be obtained only when the legitimate receiver's channel is better than the eavesdropper's channel.
When considering fading channels, main and eavesdropper's channels for a specific fading realization can be regarded as complex additive white Gaussian noise (AWGN) channels. The Shannon capacities of one realization of the quasi-static fading channels are given by
\begin{equation}\label{}
  C_M=\log_2(1+\gamma_M),
\end{equation}
\begin{equation}\label{}
  C_E=\log_2(1+\gamma_E),
\end{equation}
where $\gamma_M$ and $\gamma_E$ are the instantaneous SNRs at the legitimate receiver and the eavesdropper, respectively.
The instantaneous SNR at the legitimate receiver is equal to $\gamma_M=P|h_M|^2/\sigma^2_M$, where $P$ denotes the transmit power, $h_M$ denotes the instantaneous channel gain at the legitimate receiver, and $\sigma^2_M$ denotes the receiver noise variance at the legitimate receiver.
Also, the instantaneous SNR at the eavesdropper is equal to $\gamma_E=P|h_E|^2/\sigma^2_E$, where $h_E$ denotes the instantaneous channel gain at the eavesdropper and $\sigma^2_E$ denotes the receiver noise variance at the eavesdropper.
Thus, the secrecy capacity for one realization of the quasi-static fading channels can be written as
\begin{flalign}\label{eq:Cs_one}
C_S=\begin{cases}
  \log_2(1+\gamma_M)-\log_2(1+\gamma_E), \quad \text{if} \quad \gamma_M>\gamma_E,\\
  0, ~\quad \quad \quad\quad \quad \quad \quad \quad\quad\quad\quad\quad~\text{otherwise.}
  \end{cases}
\end{flalign}
Note that in order to achieve the secrecy capacity in Eq.~(\ref{eq:Cs_one}), the transmitter needs perfect knowledge of both $\gamma_M$ and $\gamma_E$.

To measure the performance of secure transmissions over fading channels with imperfect CSI,
ergodic secrecy capacity and outage-based characterizations are often adopted.
In what follows, focusing on these two kinds of characterizations, we provide a review on the information theoretic aspect of the research in the field of physical layer security with imperfect CSI. In addition, a brief description of the secrecy  degrees of freedom, which applies for systems with
pessimistic and strong CSI assumptions, is presented as well.

\subsection{Ergodic Secrecy Capacity}\label{sec:IT_ESC}
Ergodic secrecy capacity applies for delay tolerant systems where the encoded messages are assumed to span sufficient channel realizations to capture the ergodic features of the channel. It captures the capacity limit under the constraint of perfect secrecy. Typical examples of delay tolerant applications are document transmission and e-mail, which belong to the non-real time data traffic.

Gopala et al.~\cite{Gopala_08} presented ergodic secrecy capacity for both full CSI case and the case of only main channel CSI available at the transmitter.
The secrecy capacity for one realization of the quasi-static fading channels is given in Eq.~(\ref{eq:Cs_one}).
Averaging over all fading realizations, the ergodic secrecy capacity of fading channels with full CSI is given by
\begin{equation}\label{eq:erg_F}
  \bar{C}_S^{(F)}\!\!=\!\!\int_0^{\infty}\!\!\!\int_{\gamma_E}^{\infty}\!\!\!\left(\log_2(1\!+\!\gamma_M)\!-\!\log_2(1\!+\!\gamma_E)\right)\!f(\!\gamma_M\!)f(\!\gamma_E\!)\mathrm{d}\!\gamma_{\!M}\mathrm{d}\!\gamma_{\!E}
\end{equation}
where $f(\gamma_M)$ and $f(\gamma_E)$ are the distribution functions of $\gamma_M$ and $\gamma_E$, respectively. Since the transmitter has full CSI on both channels, the transmitter can make sure that the transmission occurs only when $\gamma_M>\gamma_E$.
When only the channel gain of the legitimate receiver is known at the transmitter,
the ergodic secrecy capacity is given by
\begin{equation}\label{eq:erg_M}
  \bar{C}_S^{(M)}\!\!=\!\!\int_0^{\infty}\!\!\!\int_0^{\infty}\!\!\!\left[\log_2(1\!+\!\gamma_M)\!-\!\log_2(1\!+\!\gamma_E)\right]^+\!f(\!\gamma_M\!)f(\!\gamma_E\!)\mathrm{d}\!\gamma_{\!M}\mathrm{d}\!\gamma_{\!E}
\end{equation}
in which $[x]^+=\max\{x, 0\}$.
They outlined a variable-rate transmission scheme to show the achievability of ergodic secrecy capacity with only main channel information. During a coherence interval with the received SNR at the legitimate receiver $\gamma_M$, the transmitter transmits codewords at rate $\log_2(1+\gamma_M)$.
This variable-rate scheme relies on the assumption of large coherence intervals and ensures that when $\gamma_M<\gamma_E$, the mutual information between the source and the eavesdropper is
upper-bounded by $\log_2(1+\gamma_M)$. When $\gamma_M\ge\gamma_E$, this mutual
information is equal to $\log_2(1+\gamma_E)$. Averaging over all the fading
states, the achievable perfect secrecy rate is given by Eq.~(\ref{eq:erg_M}).
The secure message is hidden across different fading states.

\begin{figure}[!htb]
\centering\vspace{-2mm}
\includegraphics[width=1\columnwidth]{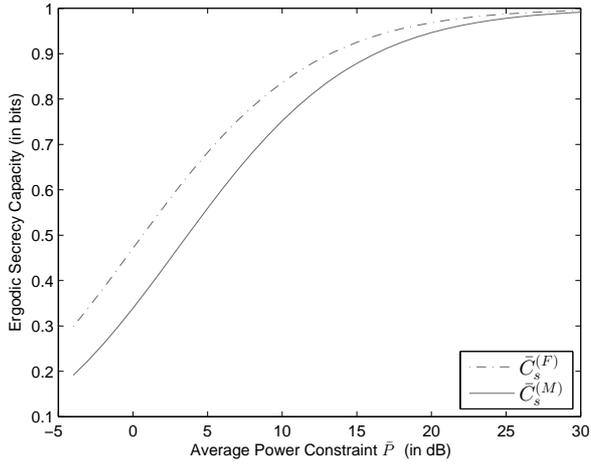}
\vspace{-6mm} \caption{Ergodic secrecy capacity versus average power constraint.  The average channel qualities are $E\left\{|h_M|^2\right\}=E\left\{|h_E|^2\right\}=1$.} \label{fig:Fig2}
\end{figure}
Fig.~\ref{fig:Fig2} compares the ergodic secrecy capacity of the network with full CSI at the transmitter and the ergodic secrecy capacity of the network with only main channel CSI at the transmitter. The average channel qualities are $E\left\{|h_M|^2\right\}=E\left\{|h_E|^2\right\}=1$, where $E\{\cdot\}$ is the expectation operation. The average power constraint is denoted by $\bar{P}=E\{P\}$.
According to Eq.~(\ref{eq:erg_F}), the transmission occurs only when  $|h_M|^2>|h_E|^2$. Thus, the constant power level used for transmission with full CSI at the transmitter is $P=\bar{P}/\Pr(|h_M|^2>|h_E|^2)$, where $\Pr(\cdot)$ denotes the probability measure. Note that $P=0$ for $|h_M|^2\le |h_E|^2$, and hence $E\{P\}=\bar{P}$.
On the other hand, the constant power level used for transmission with only main channel CSI at the transmitter is $P=\bar{P}$.

In addition, Khisti and Wornell studied the ergodic secrecy capacity for multiple-input, single-output, multiple-eavesdropper (MISOME) systems in~\cite{Khisti_10}. They developed upper and lower bounds on the ergodic secrecy capacity with perfect CSI on legitimate receiver's channel and imperfect CSI on eavesdropper's channel. They also investigated the ergodic secrecy capacity of fasting fading channel for both high SNR and finitely many antennas, i.e, the number of transmitted antenna is very large.
Rezki et al.\cite{Rezki_11,Rezki_12} studied the ergodic secrecy capacity for systems with imperfect CSI on both legitimate receiver's and eavesdropper's channels at the transmitter. In~\cite{Rezki_11}, they presented a framework that characterizes the ergodic secrecy capacity of fast fading channels under imperfect legitimate receiver's channel estimation at the transmitter. In~\cite{Rezki_12}, they established upper and lower bounds on the ergodic secrecy capacity for single-input, single-output, single-eavesdropper (SISOSE) system with limited-rate feedback of the legitimate receiver's channel information.

\subsection{Outage-Based Characterizations}
As mentioned before, the ergodic secrecy capacity applies for delay tolerant systems which allow for the adoption of an ergodic version of fading channels. However, perfect secrecy cannot always be achieved for systems with stringent delay constraints, and ergodic secrecy capacity is inappropriate to characterize the performance limits for such systems. On the other hand, outage-based characterizations, which measure systems with probabilistic formulations, become more appropriate.

In~\cite{Parada_05}, assuming the fading is quasi-static, Parada and Blahut analyzed the scenario where the CSI of both legitimate receiver's and eavesdropper's channels is not available at the transmitter. They provided an alternative definition of outage probability, wherein secure communications can be guaranteed for the fraction of time when legitimate receiver's channel is stronger than eavesdropper's channel.
Barros and Rodrigues~\cite{Barros_06} firstly provided a detailed characterization of the outage secrecy capacity where the outage probability, $p_{\text{out}}$, is characterized by the probability that a given target rate, $R_S$, is greater than the difference between instantaneous main channel capacity, $C_M$, and instantaneous eavesdropper's channel capacity, $C_E$. The expression of $p_{\text{out}}$ is given by
\begin{equation}\label{eq:pout}
  p_{\text{out}}=\Pr\left(C_M-C_E<R_S\right).
\end{equation}
They also showed that fading alone guarantees that physical layer security is achievable, even when the eavesdropper has a better average SNR than the legitimate receiver.
In addition, Bloch et al. characterized the relationship between the upper bound of outage probability and the variance of the channel estimation error on eavesdropper's channel in~\cite{Bloch_08}.
The secrecy outage behavior of a multiple-input, single-output, single-eavesdropper (MISOSE) fading system is studied in~\cite{Gerbracht_12}, where the authors proposed a relation between the degree of channel knowledge and the tolerable secrecy outage probability. 

In~\cite{Parada_05,Barros_06,Bloch_08,Gerbracht_12}, the outage-based formulations capture the probability of having a reliable and secure transmission. Reliability and security are not distinguished, since an outage occurs whenever the transmission is either unreliable or not perfectly secure.
In~\cite{Zhou_11}, Zhou et al. presented an alternative secrecy outage formulation which directly measures the probability that a transmitted message fails to achieve perfect secrecy.
The alternative secrecy outage is formulated by
\begin{equation}\label{eq:pso}
  p_{\text{so}}\!=\!\Pr\left(C_E\!>\!R_M-R_S \!\mid\! \text{message transmission}\right),
\end{equation}
where $R_M$ and $R_S$ are the rate of transmitted codeword and the rate of the confidential information in the wire-tap code, respectively, and the outage probability is conditioned on a message actually being transmitted.
From Eq.~(\ref{eq:pso}), we see that the new formulation takes into account the system design parameters, such as the rate of transmitted codewords and the condition under which message transmissions take place. Therefore, the alternative secrecy outage formulation is useful for the system designer to design transmission schemes that meet target security requirements.

\begin{figure}[!htb]
\centering\vspace{-2mm}
\includegraphics[width=1\columnwidth]{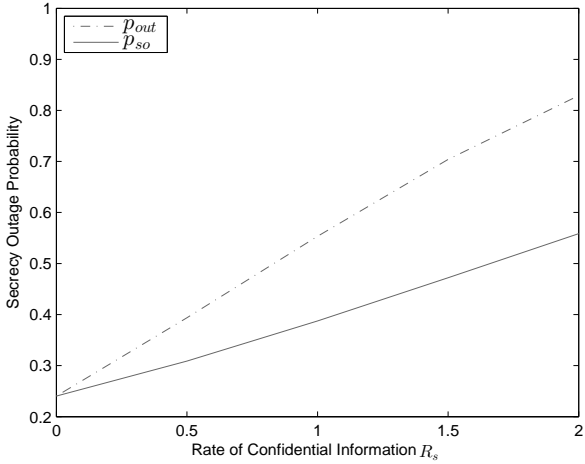}
\vspace{-6mm} \caption{Secrecy outage probability versus rate of confidential information. The average received SNRs are $E\{\vr_M\}=E\{\vr_E\}=1$.} \label{fig:Fig3}
\end{figure}
Fig.~\ref{fig:Fig3} compares the secrecy outage probability of not having a reliable and secure transmission, $p_{out}$ in Eq.~(\ref{eq:pout}), and the secrecy outage probability of not having a secure transmission, $p_{so}$ in Eq.~(\ref{eq:pso}). The average received SNRs are $E\{\vr_M\}=E\{\vr_E\}=1$.
The detailed discussion on the comparison of these secrecy outage probabilities can be found in~\cite{Zhou_11}.

\subsection{Secrecy Degrees of Freedom}
Apart from ergodic secrecy capacity and outage-based characterizations, another line of research, e.g.,~\cite{He_10,He_11_1,He_11_2,He_11_3,He_12}, studied the performance limits of systems with imperfect CSI under a pessimistic but strong assumption that allows the eavesdropper's channel to be arbitrarily varying. These work analyzed the so-called secrecy degrees of freedom (s.d.o.f), which is the pre-log of the secrecy capacity at high SNR and captures the asymptotic behavior of the achievable secrecy rate in high SNR regime.
The s.d.o.f. is formulated as
\begin{equation}\label{}
  \text{s.d.o.f.}=\lim_{\bar{P}\rightarrow\infty}\sup\frac{R_S}{\log_2\left(\bar{P}\right)},
\end{equation}
where $\bar{P}$ denotes the average power constraint on transmitted signals.

The s.d.o.f region for single-user Gaussian multi-input, multi-output (MIMO) wiretap channel was investigated in~\cite{He_10}.
The s.d.o.f. region of two user Gaussian MIMO broadcast channel with an arbitrarily varying eavesdropper channel was found in~\cite{He_11_1}. The s.d.o.f. region for two user Gaussian MIMO multiple access channel and Gaussian two-way channel with the eavesdropper channel being arbitrarily varying were given in~\cite{He_11_2} and~\cite{He_11_3}, respectively.
The work of~\cite{He_12} studied s.d.o.f. region for two-user MIMO interference channel with an external eavesdropper.
In addition, there is a main limitation of this type of works. It always requires some advantage in the antenna numbers on the legitimate receiver over the eavesdropper to get positive s.d.o.f..


\section{Signal Processing Secrecy Enhancements}\label{sec:SP}
In this section, we present various signal processing techniques for enhancing secrecy of wireless communications. Specifically, secure on-off transmissions for signal-antenna channels, beamforming with artificial noise for multi-antenna channels, and secure design techniques for relay channel and cognitive radio systems are described in the following three subsections.

\subsection{Secure On-off Transmissions for Single-Antenna Channels}
Secure on-off transmission policy in wireless network designs generally works in the following way. The transmitter decides whether or not to transmit according to the knowledge of CSI on the legitimate receiver's channel or eavesdropper's channel or both channels (if applicable). Transmissions take place whenever the estimated instantaneous CSI fulfills the requirements related to some given thresholds, e.g., SNR thresholds. Otherwise, transmissions are suspended.

Gopala et al.~\cite{Gopala_08} proposed a low-complexity on-off power allocation strategy according to the instantaneous CSI on the legitimate receiver's channel, which approaches the optimal performance for asymptotically large average SNR.
Zhou et al.~\cite{Zhou_11} designed two on-off transmission schemes, each of which guarantees a certain level of security whilst maximizing the throughput. With the statistics of eavesdropper's channel information, the first scheme requires CSI feedback from the legitimate receiver to the transmitter, and the second scheme requires only 1-bit feedback.
In~\cite{Rezki_11}, Rezki et al. studied the system with imperfect legitimate receiver's CSI and statistics of eavesdropper's channel at the transmitter. They derived the achievable rate of fast fading channel with a simple on-off scheme using a Gaussian input.
In addition, under various assumptions on the CSI, He and Zhou~\cite{He_13_2} proposed several secure on-off  transmission schemes, which maximize the throughput subject to a constraint on secrecy outage probability.  Both fixed-rate and variable-rate transmissions were proposed. Also, they not only considered the imperfect CSI at the transmitter, but also studied the impact of imperfect CSI at the receiver side.



\subsection{Beamforming with Artificial Noise for Multi-Antenna Channels}
\begin{figure}[!htb]
\centering\vspace{-2mm}
\includegraphics[width=1\columnwidth]{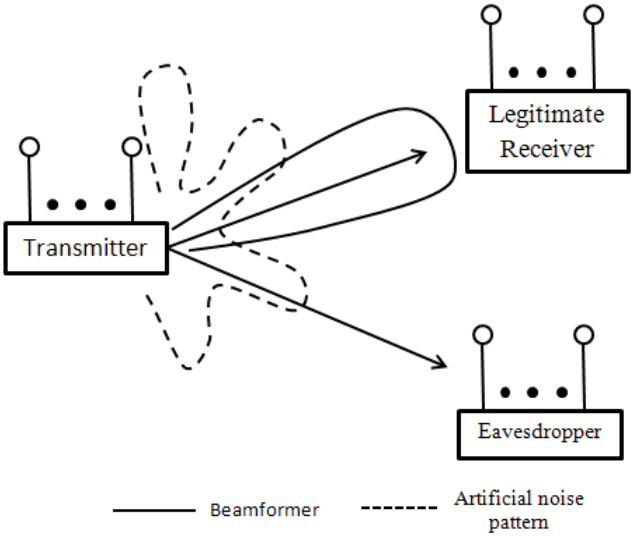}
\vspace{-6mm} \caption{An illustration of beamforming with artificial noise.} \label{fig:Fig4}
\end{figure}

The work by Hero~\cite{Hero_03} is arguably the first to consider secret communication in a multi-antenna transmission system, and sparked significant efforts to this problem~\cite{Mukherjee_10}.
For multi-antenna channels with imperfect CSI, beamforming with artificial noise  is the one of the most widely-used techniques to secure the data transmission.
Negi and Goel~\cite{Negi_05,Goel_08} first proposed an artificial noise injection strategy. In addition to transmit information signals, part of the transmission power is allocated to generate artificial noise in order to confuse the eavesdropper. Specifically, the produced artificial noise lies in the null space of the legitimate receiver's channel, while the information signal is transmitted in the range space of the legitimate receiver's channel. This technique relies on the instantaneous CSI on legitimate receiver's channel, but does not require the instantaneous CSI on eavesdropper's channel. The legitimate receiver's channel nulls out the artificial noise. Thus the legitimate receiver is not affected by the noise.
The basic idea of beamforming with artificial noise is presented in Fig.~\ref{fig:Fig4}.

In the following, we illustrate the work after Negi and Goel's research, which studied the beamforming with artificial noise in various multi-antenna channel scenarios with different assumptions on the availability of CSI. We first present the literature considering imperfect CSI on the eavesdropper's channel, and then discuss the work which considers the imperfect CSI on both eavesdropper's and legitimate receiver's channels.

\subsubsection{Imperfect CSI on Eavesdropper's Channel}\label{sec:MDIm}
The work in~\cite{Zhou_10} studied the optimal power allocation
between the information signal and the
artificial noise for systems with both non-colluding and colluding eavesdroppers.
The authors found that the equal power allocation is the strategy that achieves nearly the same secrecy rate as the optimal power allocation for the non-colluding eavesdropper case and more power should be used to transmit the artificial noise as the number of eavesdropper increases for the colluding eavesdropper case.
In~\cite{Zhang_13}, Zhang et al. investigated the design of artificial-noise-aided secure multi-antenna transmission in slow fading channels. They provided throughput-maximizing design solutions
with either fixed-rate or adaptive-rate encoder, both including the optimal rate parameters of
the wiretap code as well as the wise transmit power allocation between the information signal and the artificial noise.
Huang and Swindlehurst~\cite{Huang_12} obtained the robust transmit covariance matrices on worst-case secrecy rate maximization under both individual and global power constraints.
They investigated both cases of direct transmission and cooperative jamming with a helper.
In addition, Gerbracht et al.~\cite{Gerbracht_12} characterized the optimal single-stream beamforming with the use of artificial noise to minimize the outage probability. They pointed out that the solution converges to maximum ratio transmission (MRT) for the case of no CSI to the eavesdropper, and the optimal beamforming vector converges to the generalized eigenvector solution with the growing level of CSI.
Lin et al.~\cite{Lin_12} showed that the artificial noise selected in~\cite{Goel_08} is suboptimal.
According to their study, the eigenvectors of the optimal covariance matrices of both
information signals and generalized artificial noise are equal to the right singular vectors of the legitimate receiver's channel, and the power of artificial noise should be allocated uniformly over the eigenvectors. The rigorous proofs for these facts were also provided.


In the work of~\cite{Zhou_10,Zhang_13,Huang_12,Gerbracht_12,Lin_12,Goel_08}, although the instantaneous CSI on the eavesdropper's channel is not required, the transmitter still needs the statistics of eavesdropper's channel. For the case where no CSI on the eavesdropper's channel (including the statistics) is known at the transmitter, Swindlehurst and Mukherjee~\cite{Swindlehurst_09,Mukherjee_09} proposed a modified water-filling algorithm which balances the required transmit power with the number of
spatial dimensions available for jamming the eavesdropper. As described in the modified water-filling algorithm, the transmitter first allocates enough power to meet a target performance criterion, e.g., SNR or rate, at the receiver, and then use the remaining power for broadcasting artificial noise.
They also applied the similar algorithm to investigate the multiuser downlink channels in~\cite{Mukherjee_09_2}.

\subsubsection{Imperfect CSI on both Eavesdropper's and Legitimate Receiver's Channels}
The imperfect CSI on the legitimate receiver's channel at the transmitter mainly incurs two problems. First, without knowing the actual instantaneous CSI on the legitimate receiver's channel, the transmitter cannot make sure that the data transmission rate is not larger than the legitimate receiver's channel capacity. Then, a transmitted packet is unable to be decoded by the receiver, i.e., the packet is corrupted, whenever the data transmission rate exceeds the legitimate receiver's channel capacity. Second, with imperfect instantaneous CSI on the legitimate receiver's channel, the artificial noise leaks into the legitimate receiver's channel, since the beamforming with artificial noise is designed according to the estimated instantaneous CSI rather than the actual instantaneous CSI. Therefore, the artificial noise interferes the desired user, despite that it is intended to only confuse the eavesdropper. Naturally, the techniques applying for systems with perfect CSI on the legitimate receiver's channel become not optimal.


Taylor et al. presented the impact of the legitimate receiver's channel estimation error on the performance
of an eigenvector-based jamming technique in \cite{Taylor_11}. Their research showed that the ergodic secrecy rate provided
by the jamming technique decreases rapidly as the channel estimation error increases.
Mukherjee and Swindlehurst~\cite{Mukherjee_11} also pointed out that the security provided by beamforming approaches is quite sensitive to the imprecise channel estimates. They proposed a robust beamforming scheme for MIMO secure transmission systems with imperfect CSI of the legitimate receiver.
Pei et al.~\cite{Pei_12} addressed a stochastic time-varying CSI uncertainty model with uplink-downlink reciprocity.
Using this model, they proposed a new iterative algorithm to secure the transmission, which is robust to CSI errors.
The authors of~\cite{Zhou_10} investigated the effects of imperfect CSI on the
optimal power allocation and the critical SNR for secure
communications. They found that allocating power on the artificial noise to confuse the eavesdropper is better than
increasing the signal strength for the legitimate receiver as
the channel estimation error increases.
Adapting the secrecy beamforming scheme, Liu et al. \cite{Liu_12} investigated the joint design of training and data transmission signals for wiretap channels.  The ergodic secrecy rate for systems with imperfect channel estimations at both the legitimate receiver and the eavesdropper was derived.
The secrecy rate is hard to be calculated, since the channel estimation errors cause non-Gaussianity of equivalent noise. They solved this problem by the large number of transmit-antenna analysis. Based on the achievable ergodic secrecy rate, they found the optimal tradeoff between the power
used for training and data signals.
Furthermore, advocating the joint optimization of the transmit weights and artificial noise spatial distribution with the use of a quality-of-service (QoS)-based perspective, Liao et al.~\cite{Liao_11}
proposed a secret-transmit beamforming approach in accordance with the imperfect CSI on both the legitimate receiver's and eavesdropper's channels.
In~\cite{Ng_11}, Ng et al. addressed a resource-allocation and scheduling optimization problem
for orthogonal frequency division multiple access (OFDMA) networks. The optimization problem takes into account artificial noise generation and the effects of imperfect CSI in slow fading. They proposed a resource allocation algorithm considering secrecy outage, channel outage, and the potentially detrimental effect of artificial noise generation.
Considering the systems with partial CSI feedback, Lin et al. investigated the scenario where only quantized channel direction information (CDI) of the legitimate receiver's channel is available at the transmitter in~\cite{Lin_11}.
For a given transmission power and a fixed number of feedback bits, they derived the optimal power allocation among the information signal and the artificial noise to maximize the secrecy rate under artificial noise leakage.

\vspace{1mm}
In addition, it is necessary to mention that there also exist some studies not considering the artificial noise to enhance the security of multi-antenna systems with imperfect CSI. Some examples, i.e.,~\cite{Li_11_J,li_11,Lin_12,Geraci_13}, of these studies are illustrated as follows.
In~\cite{Li_11_J}, Li and Petropulu solved optimal input covariance that maximizes the ergodic secrecy rate subject to a power constraint for MISOSE systems with imperfect CSI on eavesdropper's channel.
Li and Ma~\cite{li_11} formulated a transmit-covariance optimization problem for secrecy-rate
maximization (SRM) of MISOME systems with imperfect CSI on both main and eavesdropper's channels.
The authors of~\cite{Lin_12} analyzed the systems with only statistics of main and eavesdropper's channels at the transmitter. They showed that the secure beamforming is still secrecy capacity achieving in such a scenario, and proposed the optimal channel input covariance matrix, which fully characterizes the secrecy capacity. They also pointed out that the artificial noise is not necessary in this case.
Geraci et al. studied secrecy sum-rates achievable by regularized channel inversion (RCI) precoding in MISO systems under imperfect CSI in~\cite{Geraci_13}.

\subsection{Secure Designs for Relay Channels and Cognitive Radio Systems}
Secure communication assisted by relay nodes is often regarded as a natural extension to the secure transmission in multi-antenna networks. The physical layer security can be provided by careful signaling at different relays in the system.
A virtual beam towards the legitimate receiver can be built by the collaboratively work among relay nodes, which is similar to the secure transmission in multi-antenna systems.
However, unlike the multiple-antenna transmission, the transmitter cannot directly control the relays.
For the network of single-antenna wiretap channel with serval relays,
Goel and Negi~\cite{Goel_08} described a 2-phase protocol to obtain coordination in transmitting artificial noise among the relays.
In the first phase, the transmitter and the legitimate receiver both transmit independent artificial noise signals to the relays. Different linear combinations of these two signals are received by the relays and the eavesdropper.
In the second phase, the relays replay a weighted version of the received signal, using a publicly
available sequence of weights. Meanwhile, the transmitter transmits the confidential information, along with a
weighted version of the artificial noise transmitted in the first stage.
With the knowledge of the artificial noise component due to the legitimate receiver, the legitimate receiver is able to cancel off the artificial noise and get the confidential information.
Assuming global full CSI at every node, the work in~\cite{Dong_10} provided a detailed analysis on secure communications
of one source-destination pair with the help of multiple cooperating relays in the presence of one or more eavesdroppers, which includes decode-and-forward (DF), amplify-and-forward (AF), and cooperative jamming (CJ) three different cooperate schemes.

To explore the effects of imperfect CSI in relay systems, researchers often consider the uncertainty of CSI on three kinds of links, which are relay-destination links, relay-eavesdropper links, and source-relay links.
The authors of~\cite{Dong_08} investigated the effect of imperfect CSI on the relay-eavesdropper channels. They proposed a DF relaying protocols for secure communication, which maximizes the lower bound on the ergodic secrecy capacity under a total relay transmission power constraint.
Considering the imperfect CSI on the channels from relay to destination and relay to eavesdropper, Zhang and Gursoy~\cite{Zhang_10} provided optimization frameworks for the robust DF-based relay beamforming design.
Furthermore, Vishwakarma and Chockalingam~\cite{Vishwakarma_12} computed the worst case secrecy
rate when there are imperfections in the CSI on all the links, i.e., relay-destination links, relay-eavesdropper links, and source-relay links.

Cognitive Radio (CR) has been widely recognized as an effective technology to improve the utilization of wireless spectrum by allowing secondary users to coexist with primary users and access the spectrum of the prime system.
In CR systems with secrecy message broadcasted in the primary links, the signals of secondary links for their users can also serve as the artificial noise for the secure primary link.
In order to confuse the eavesdropper overhearing the primary links,
the secondary system operates similar to that of the helper nodes, but simultaneously serving their own receivers.
The papers that study the imperfect CSI in such cognitive radio systems can be found in~\cite{Kwon_12} and references within.  The authors of \cite{Kwon_12} explored MISO CR systems where the secondary system secures the primary communication in
return for permission to use the spectrum.
On the other hand, when the secondary user transmitter sends confidential information to a secondary user receiver on the same frequency band with a primary user, the requirement of not interfering the primary users is often treated as a power constraint on the transmitted signals in the secondary system.
Assuming all CSI is imperfect known, Pei et. al~\cite{Pei_11} explored the optimal secondary user transmitter design, which maximizes the secure transmission rate of the secondary link while avoiding harmful interference to the primary users.
They proposed two approaches to solve this challenging optimization problem, which is non-convex and semi-infinite.

%
\section{Secrecy in Large-Scale Decentralized Wireless Networks}\label{sec:LargeSc}

In the last section, we summarized the research results on physical layer security enhancements for systems consisting of a small number of nodes. We now turn our attention to another very important class of wireless networks: large-scale decentralized wireless networks. In such networks, the CSI of eavesdroppers is rarely available at legitimate users. Even the locations of eavesdroppers may not be known. The lack of eavesdropper's information makes communication security a challenging problem. Apart from that, the decentralized nature of the network rules out any global optimization approach for secrecy enhancements. Pioneering works on physical layer security in large-scale decentralized wireless networks focused on the connectivity analysis. Specifically, the notion of secrecy graph was introduced in~\cite{Haenggi08} and further developed in~\cite{Pinto12a} to include fading channels. Various secure connectivity improvements were discussed in~\cite{Zhou11a,Pinto12b}, including multi-antenna sectoring and beamforming. The connectivity in the presence of the location uncertainty of eavesdroppers was studied in~\cite{Goel10}.

Building up from the connectivity analysis of the secrecy graph, the secrecy capacity scaling was analyzed in~\cite{Vasudevan10,Koyluoglu12,Liang09,Capar12a,Capar12b}. Specifically, the studies in~\cite{Vasudevan10,Koyluoglu12} showed that the secrecy requirement does not reduce the capacity scaling of the network, i.e., the capacity scaling law is the same for both insecure message transmission and secure message transmission. Of course, achieving such an optimal scaling law under the secrecy constraint requires very different transmission and access protocols. For example, when eavesdroppers' locations are unknown, various secrecy enhancements such as cooperative jamming and multi-path transmission in conjunction with network coding may be required~\cite{Capar12a,Capar12b}.

Although the scaling law results may provide insights into the asymptotic secrecy throughput performance of large-scale networks, a finer view of throughput is necessary to better understand the impact of key system parameters and transmission protocols, since most of design choices affect the actual (non-asymptotic) throughput but not the scaling behaviors. To this end, a new performance metric named secrecy transmission capacity was developed in~\cite{Zhou11b,Zhou12} to capture the area spectral efficiency of secure transmission. The formulation of such a metric was based on the outage approach in~\cite{Zhou_11} which accommodates the practical scenario where the CSI of the eavesdroppers is unknown to the legitimate nodes.

\section{Open Problems and Discussions}\label{sec:OP}
Despite the increasing attention paid to the effect of imperfect CSI on the physical layer security, research on this area is still at an early stage.
In this section, we discuss some open problems in the research area of the physical layer security with imperfect CSI.

\subsection{Imperfect Channel Estimation at Receivers}
Among the existing work on physical layer security considering imperfect CSI, most investigates the impact of imperfect CSI at the transmitter while assuming perfect channel estimation at the receivers. However, only a limited amount of work, e.g., \cite{Zhou_10,Liu_12,He_13_2}, paid attention to the imperfect channel estimation at the receivers.
Clearly, the assumption of perfect channel estimation at the receiver is not very practical, since the estimation of fading channels generally is not error-free.
In principle, the channel estimation error exists at both the legitimate receiver and the
eavesdropper. Assuming perfect estimation at the eavesdropper is more reasonable from the secure transmission design point of view, since it is often difficult or impossible for the transmitter to know the accuracy of the eavesdropper's channel estimate. Nevertheless, in scenarios where the eavesdropper is just an ordinary user of the network whose performance and other information can be tracked by the transmitter, e.g., \cite{Liang_08,Khisti_10,Leow_11}, the consideration of imperfect channel estimation at the eavesdropper becomes relevant. 

\subsection{Imperfect Knowledge of Eavesdroppers' Locations}
With few exceptions, almost all the existing research studies on physical layer security assumed that the eavesdropper's location is perfectly known. The validity of such an assumption strongly depends on the application under investigation.
For example, when the eavesdropper is a passive entity without transmission, its location is very hard to be obtained. Also, in large-scale complex networks, it is very difficult to obtain eavesdroppers' locations due to the random deployments or mobility of nodes.
Therefore, it is an interesting research direction to consider the secrecy in networks with imperfect or no knowledge of the eavesdropper's location at the transmitter.


%
%
%


\section{Conclusions}\label{sec:Conc}
In this paper, we reviewed the research in physical layer security with practical assumptions on fading channel information. For the characterizations of performance limits, we described ergodic secrecy capacity suitable for delay tolerant systems, outage-based characterizations for systems with stringent delay constraints, and secrecy degrees of freedom. Also, we surveyed the signal processing secrecy enhancements proposed for different transmission scenarios, i.e., secure on-off transmission for signal-antenna channels, beamforming with artificial noise for multi-antenna channels, and other signaling designs for relay channels or cognitive radio systems. In addition, the recent results on secrecy in large-scale decentralized wireless networks were reviewed.
Future research directions on physical layer security with imperfect CSI include the imperfect channel estimation at receiver sides and the imperfect knowledge of the eavesdropper's location.

\bibliographystyle{IEEEtran}

\end{document}